\documentclass[12pt,preprint]{aastex}
\usepackage{amsmath}
\usepackage{graphicx}
\usepackage{epsfig}
\usepackage{url}
\usepackage{indentfirst}
\usepackage{color}

\newcommand{\psr}{PSR~J2021+4026}
\begin{document}

\title{Mode change of a gamma-ray pulsar, PSR~J2021+4026}
\author{J.~Zhao\altaffilmark{1}, C.W.~Ng\altaffilmark{2}, L.C.C.~Lin\altaffilmark{3}, J.~Takata\altaffilmark{1},
  Y.~Cai\altaffilmark{1}, C.-P.~Hu\altaffilmark{2}, D.C.C. Yen\altaffilmark{4}, P.H.T.~TAM\altaffilmark{5},
  C.Y.~Hui\altaffilmark{6}, A.K.H.~Kong\altaffilmark{7,8}, K.S.~Cheng\altaffilmark{2}
}
  \email{M201570114@hust.edu.cn, takata@hust.edu.cn}
 \altaffiltext{1}{School of Physics, Huazhong University of Science and Technology, Wuhan, 430074, China}
 \altaffiltext{2}{Department of Physics, The University of Hong Kong, Pokfulam Road, Hong Kong}
\altaffiltext{3}{Academia Sinica, Institute of Astronomy and Astrophysics, Taipei, 10617, Taiwan}
\altaffiltext{4}{Department of Mathematics, Fu Jen Catholic University, New Taipei City,  24205, Taiwan}
\altaffiltext{5}{Institute of Astronomy and Space Science, Sun Yat-Sen University, Zhuhai, 519082, China}
\altaffiltext{6}{Department of Astronomy and Space Science, Chungnam National University, Daejeon 305-764,
  Republic of Korea}
\altaffiltext{7}{Institute of Astronomy and Department of Physics, National
Tsing Hua University, Hsinchu, Taiwan}
\altaffiltext{8}{Astrophysics, Department of
  Physics, University of Oxford, Keble Road, Oxford OX1 3RH, UK}

\begin{abstract}
  A glitch of a pulsar is known as a sudden increase in the spin frequency and spin-down rate (frequency time derivative), and it can be caused by a sudden release of the stress built up in the solid crust of the star or pinned vortices in the superfluid interior. 
  PSR J2021+4026 is the first pulsar that shows a significant change in the gamma-ray flux and pulse profile at the glitch that occurred around 2011
  October 16.
  We report the results of timing and spectral analysis of PSR~J2021+4026 using $\sim$ 8~yr Fermi-LAT data.  
  We find that the pulsar stayed at a high spin-down rate ($\sim 4\%$ higher than the pre-glitch value) and a low gamma-ray state ($\sim 18\%$ lower) for about 3~yr after the glitch. 
  Around 2014 December, the spin-down rate and gamma-ray
  flux gradually returned to pre-glitch values
  within a time scale of a few months. 
  The phase-resolved spectra and pulse profiles after the relaxation are also consistent with those before the glitch. 
  The observed long-term evolution of the spin-down rate and the gamma-ray flux indicates that the glitch triggered a mode change in the
  global  magnetosphere.
  We speculate that the glitch changed the local magnetic field structure around the polar cap and/or the inclination angle of the dipole axis, leading to
  a change in the electric current circulating in the magnetosphere.

\end{abstract}


\section{Introduction}

The \emph{Fermi} Large Area Telescope (hereafter \emph{Fermi}-LAT) has found pulsed gamma-ray emission from more than 200 pulsars\footnote{https://confluence.slac.stanford.edu/display/GLAMCOG/Public+List+of+LAT-Detected+Gamma-Ray+Pulsars}. Among them, PSR~J2021+4026 is the only one that exhibits a mode change in the observed gamma-rays.
Its gamma-ray detection was first discovered by \emph{Fermi}-LAT in the first year of the \emph{Fermi} mission \citep{abdo09a} as a bright gamma-ray source (0FGL J2021.5+4026) in the Cygnus region of the  Milky Way.
A subsequent frequency search of the LAT data performed by \citet{abdo09b} confirmed that 0FGL J2021.5+4026 is a young pulsar with a spin period $P$ = 265ms
and a spin-down rate of $\dot{P}= 5.48\times 10^{-14} {\rm s~s^{-1}}$.
The spin-down behavior infers a luminosity of $E_{sd}\sim 10^{35}{\rm erg~s^{-1}}$, a characteristic age of $\tau_c\sim77$~kyr, and a surface dipole field of $B_d\sim 4\times 10^{12}$G.
An X-ray counterpart (2XMMJ202131.0+402645) was identified by XMM-Newton \citep{trepl10}, and its X-ray emission
was interpreted as emission from a hot spot on the surface (see \citealt{Lin&Hui, Lin16} for  detail).

  Allafort et al. (2013, hereafter A13) reported a sudden drop in the gamma-ray flux of PSR J2021+4026. They found that a glitch occurred near MJD 55,850 (2011 October 16) with a timescale smaller than a week, and it increased the spin-down rate from $|\dot{f}|=(7.8\pm 0.1)\times 10^{-13}{\rm Hz~s^{-1}}$ to $(8.1\pm 0.1)\times 10^{-13}{\rm Hz~s^{-1}}$.
At the glitch, (1) the flux ($>100$MeV) was decreased by $\sim 18$\%, from $(8.33\pm0.08)\times 10^{-10}{\rm erg~cm^{-2}~s^{-1}}$ to $(6.86\pm0.13)\times 10^{-10}{\rm erg~cm^{-2}~s^{-1}}$, 
(2) the pulse profile was changed significantly ($>5\sigma$), and 
(3) the gamma-ray  spectrum experienced a marginal change $(<3\sigma)$. 
Before the glitch, the pulse profile consisted of two strong peaks plus a small bump (probably a third peak) 
in the bridge region (BR). After the glitch, there was no evidence of the third peak, and the pulse profile was well fit by two Gaussian components.
The glitch event caused the flux drop for all rotation phases, but the strongest main peak (the second peak in A13) only had a minor change in
intensity compared to other pulse phases. 

 The \emph{Fermi}-LAT enables us to study the detailed evolution of the gamma-ray emission after the glitch event.
 \citet{ng16} reanalyzed the long-term light curve of PSR~J2021+4026 using
 7~yr \emph{Fermi}-LAT data.
 They confirmed that the gamma-ray flux did not show a gradual recovery toward the value before the glitch, and the pulsar stayed at the low-flux stage.  However, they also noted another jump
 of the gamma-ray flux at about 3 yr (around MJD 57,000) after the
 glitch occurred near MJD 55,850.
 Although they speculated that this jump was caused by  another glitch, they
 did not analyze the timing parameter beyond MJD~56,580.

In this paper, we revisit PSR~J2021+4026 and carry out the timing and spectral analysis using \emph{Fermi}-LAT data. 
We confirm that the pulsar stayed at a high spin-down rate ($\sim 4$\% higher than the value before the glitch) for about 3 yr. 
We show that after 3 yr (around 2014 December) both the spin-down rate and the gamma-ray flux returned to pre-glitch values within a time scale of a few months, indicating that the second flux jump noted by Ng et al. (2016)
was caused by the relaxation rather than an anti-glitch of this system. 
In this article, we present the flux variability in section~\ref{fluxev}, the timing analysis in section~\ref{timing}, and pulse profile/phase-resolved spectra in section~\ref{pulsef}. 
In section~\ref{disc}, we discuss the mechanisms triggering the change of both the gamma-ray flux and the timing properties.

\section{Observations}
\subsection{Flux variability}
\label{fluxev}
In the analysis of the flux variability, we use $\sim$101 months of Fermi-LAT data from 2008 August 11 to 2016 December 31, more than 1 yr of Ng et al. (2016). 
The data (latest version P8R2) that we select from the  \emph{Fermi} website\footnote{http://fermi.gsfc.nasa.gov/ssc/} are in a $10^{\circ}$ radius region of interest (ROI) centered at the 3FGL J2021.5+4026 position (R.A., Dec)=$(305^{\circ}.386, 40^{\circ}.448)$ \citep{acero15} with energies between 100 MeV and 100 GeV.
We also collect all the events converting both in the front and the back
sections of the tracker (i.e. evtype = 3), and the event class is specified for the analysis of a point source (i.e. evtclass = 128). 
In addition, we not only select those photons with zenith angles smaller than $90^{\circ}$ to reduce the gamma-ray contamination caused by Earth albedo, but we also select those data within time intervals determined as high quality (i.e. DATA\_QUAL $>$0). 
All the aforementioned data reduction in this study was performed using the \emph{Fermi} science tools v10r0p5 package.

The light curve of PSR J2021+4026 is obtained at $E>100$MeV using the binned likelihood analysis in the $Fermi$ science tools. 
First, we construct a background emission model, including the galactic diffuse emission (gll\_iem\_v06) and the isotropic diffuse emission (iso\_P8R2\_SOURCE\_V6\_v06) circulated by the \emph{Fermi} Science Support Center and all 3FGL catalog sources \citep{acero15} within $10^{\circ}$ of the  the center of the
ROI to account for the spectral contribution.
We note that the best-fit spectra for all of the background sources were obtained with the elimination of insignificant sources ($<3\sigma$) by using the gtlike tool of the science tools for $\sim$8 yr data.
Next, we divide the entire time range into 30 day time bins. 
We fix the spectral contribution according to the background emission model and refit the flux ($>100$MeV) of the target pulsar by binned likelihood analysis.
Then we can obtain the normalization constants in each time bin by estimating the photon flux of each background source.
 
\begin{figure}
   \centering
   \epsscale{1}
   \plotone{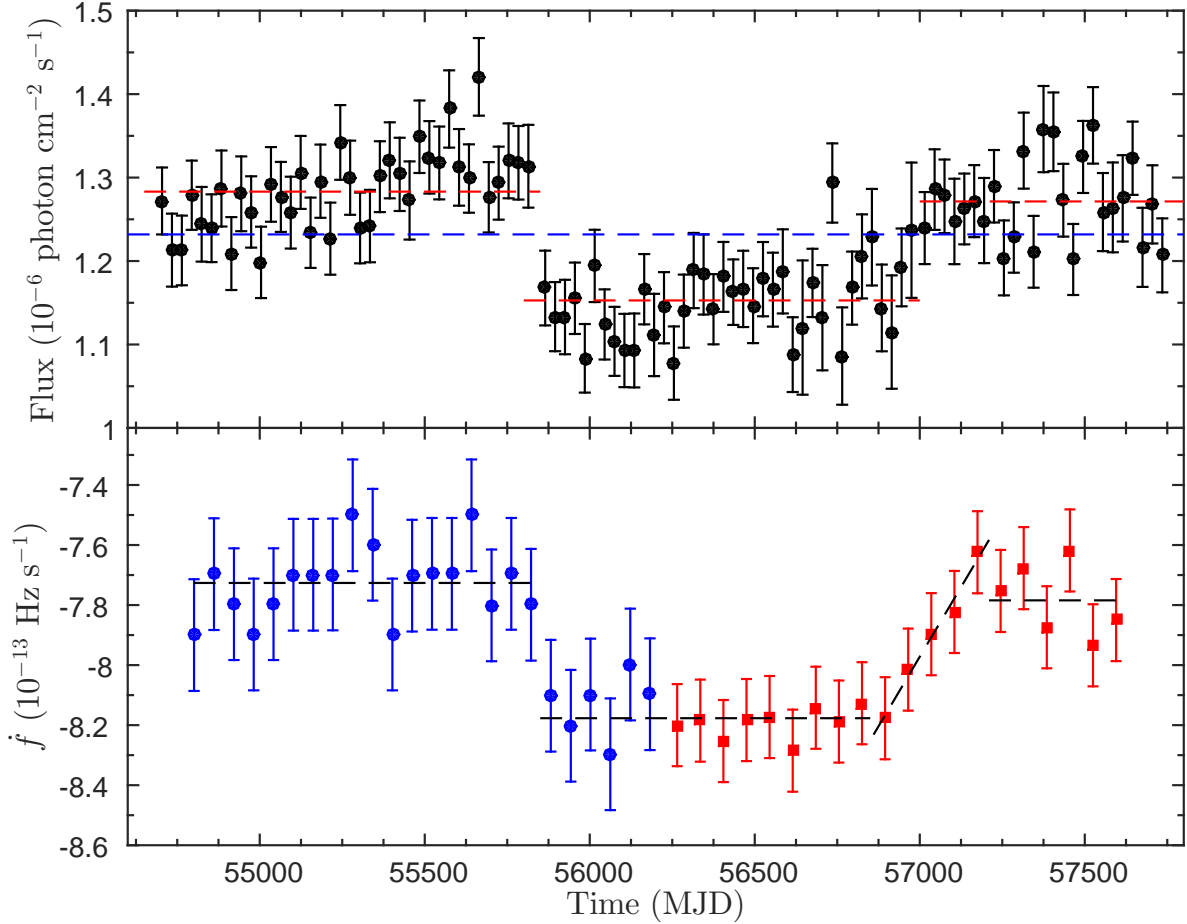}
   \caption{Evolution of the $>0.1$GeV flux (top) and the spin-down rate (bottom) of PSR J2021+4026. 
A sudden flux jump, which is accompanied by an increase in the spin-down rate, can be seen close to $\sim$MJD 55,850. 
The pulsar stayed at a lower flux and higher spin-down state until $\sim$MJD 57,000 and then gradually recovered to the pre-glitch state. 
In the bottom panel, the blue and red points are results obtained by Allafort et al.(2013)  and this work, respectively.    The uncertainty of each data point in the bottom panel denotes  the Fourier resolution/width in the
  periodicity search.   At $\sim$ MJD 56,850--57,200, an obvious increase in the spin frequency derivative can be detected for the pulsar, and a linear fit was applied to clearly indicate it. 
     Except for the epoch mentioned above, the dashed lines shown in different stages of each panel represent the weighted mean of the measured flux and the frequency derivative.}
   \label{evolution}
 \end{figure}

We obtain a time evolution of the photon flux for PSR J2021+4026 as shown in the top panel of Fig.~\ref{evolution}.
According to the flux level, the long-term light curve can be divided into three epochs/stages, which we define as 
(1) the pre-glitch stage before MJD~55,850, around which the glitch caused a sudden flux drop (A13); (2) the low-flux stage  between MJD 55,850 and 57,000; 
(3) the post-relaxation stage after MJD 57,000, around which the gamma-ray flux returned to
the value from before the glitch. 
The average fluxes for these stages (Table~1) are $F_1 = (1.29 \pm0.01) \times 10^{-6} {\rm cts~cm^{-2}~s^{-1}}$, $F_2 = (1.15\pm 0.01)\times 10^{-6} {\rm cts~cm^{-2}~s^{-1}}$, $F_3 = (1.27\pm 0.01)\times 10^{-6} {\rm cts~cm^{-2}~s^{-1}}$, respectively.
The average fluxes obtained in the pre-glitch and low-flux stages are consistent with the  results reported in A13 and \citet{ng16}. 
As shown in the figure, the flux level of the post-relaxation stage after $\sim$ MJD 57,000 does not change significantly, and it is consistent with the pre-glitch value (within $1\sigma$), suggesting that the gamma-ray emission was returned to the pre-glitch stage.

\begin{figure}
   \centering
   \epsscale{1}
   \plotone{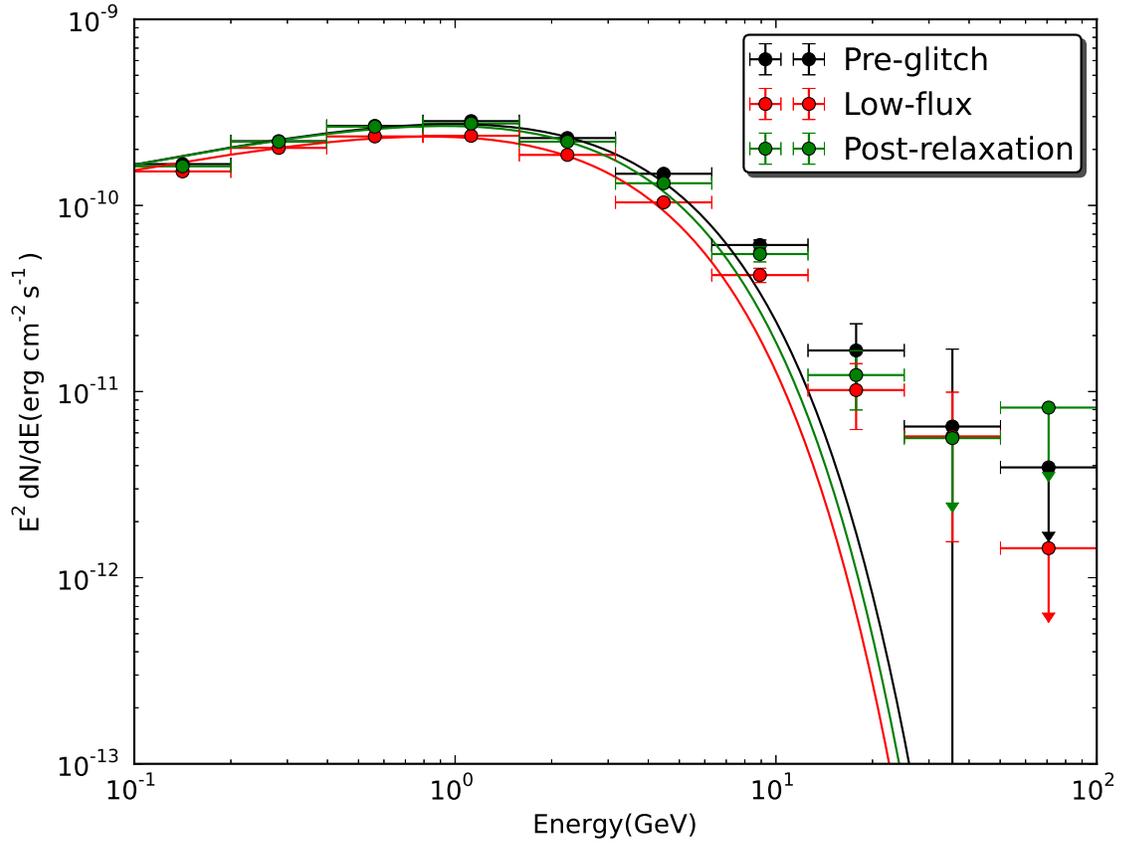}
   \caption{Phase-averaged spectra for three stages. 
   Black: Pre-glitch stage (before MJD 55,850). 
   Red: Low-flux stage (MJD 55,850--57,000). 
   Green: Post-relaxation (after MJD 57,000).
   Each spectrum is modeled by the equation~(\ref{fit}) and Table~1.}
     \label{spectrum}
 \end{figure}

The gamma-ray spectra of PSR J2021+4026 are modeled using a power
law with an exponential cutoff of the form
 \begin{equation}
   \dfrac{dN}{dE} =N_{0} \left(\frac{E}{E_{0}} \right)
   ^{\Gamma}{\rm exp}\left[-\left(\frac{E}{E_{C}} \right)^b\right],
\label{fit}
 \end{equation}
 where $N$ is the number of photons per unit time,
 unit area and unit energy; $E$ is the energy of photons; $N_{0}$ is the normalization constant; $E_{0}$ is the scale factor of energy; $\Gamma$ is the spectral power-law index; and $E_{C}$ is the cutoff energy. 
In this paper, we apply $b=1$ for both the phase-averaged and phase-resolved spectra (section~\ref{pulsef}).
Fig.~\ref{spectrum} represents the phase-averaged spectra for the three different stages, and Table~1 summarizes their best-fit cutoff energy and power-law
index.  The parameters obtained for the post-relaxation stage are also consistent with those before the glitch at a  95\% confidence level.

  \begin{table*}
    \centering
    \label{tabel1}
    \begin{tabular}{lccc}
      \hline\hline
     Parameter  &  Pre-Glitch & Low-Glux & Post-Relaxation \\\hline
     Flux, $F (10^{-6}{\rm cts~cm^{-2}~s^{-1}})$ \dotfill  & $1.29\pm 0.01$  & $1.15\pm 0.01$      & $1.27\pm 0.01$ \\ 
     Cutoff energy, $E_c$ (MeV)\dotfill & $2755\pm 74$   &  $2447\pm 77$   & $2569\pm 89$       \\ 
     Power-law  index, $\Gamma$ \dotfill& $1.640\pm 0.011$   &  $1.662\pm 0.014$       & $1.639\pm0.015$    \\ \hline
    \end{tabular}
    \caption{Parameters of phase-averaged spectra for different stages.}
  \end{table*}  
  
\subsection{Timing analysis}
\label{timing}
\psr\ is a variable gamma-ray pulsar (A13), and its ephemeris cannot be precisely described without considering the timing noise.
According to the study of the photon flux for this target (section~\ref{fluxev}), we note that the emitted gamma-ray flux gradually recovered to the pre-glitch level, and this relaxation started from $\sim$ MJD~57,000.
Since the drop in the gamma-ray emission was accompanied by  a change in the pulse profile and a sudden decrease of the frequency derivative (i.e., $\dot{f}$), the observed relaxation in the gamma-ray flux gives us a strong motivation to investigate the evolution of the timing ephemeris.
In order to investigate the change in the structure of the pulse profile and the phase-resolved spectra, precise ephemerides for different states are required.
To describe the timing behavior of \psr, we consider the ephemeris provided by the \emph{Fermi} team\footnote{http://www.slac.stanford.edu/$\sim$kerrm/fermi\_pulsar\_timing/J2021+4026/html/J2021+4026\_54683\_56587\_chol.par} and build local ephemerides of \psr\ using the \emph{Fermi}-LAT archive during $\sim$MJD~55,850--56,950 and 57,050--57,570 to cover the entire low-flux and post-relaxation stages.
More details are given in \S~\ref{sssec:ephemeris}.
We also monitor the evolution of the pulsar timing with narrower time bins in \S~\ref{sssec:parameters} to completely record a gradual recovery for the spin-down rate.

The position for all the ephemerides of \psr\ is fixed at (J2000) R.A.=$20^h21^m30^s.733$, decl.=$+40^{\circ}26'46''.04$, according to the image resolved by the  \emph{Chandra}/ACIS (Advanced CCD Imaging Spectrometer) with an uncertainty of $1''.3$ \citep{Wei2011}, and it is also the central position to extract the source events within $1^{\circ}$ radius ROI for our timing analysis.

\subsubsection{Evolution of pulsar timing}
\label{sssec:parameters}
To investigate the local evolution of timing behaviors following the change of the gamma-ray flux for \psr, we apply a method similar to that described in A13 to monitor the evolution of the pulsar timing parameters, and we extract the \emph{Fermi}-LAT data from MJD~56,580--57,630, covering a part of the low-flux and the post-relaxation stages.
All the data are divided into the 70-day bins, which have enough source photons to significantly detect the pulsation and can ignore/minimize the contamination of the high-order timing noises so that the frequency evolution can be simplified as a linear relation of $f(t) = f(t_0)+\dot{f}\times (t-t_0)$, where $f$, $\dot{f}$, and $t_0$ denote the spin frequency, frequency derivative, and epoch zero for  each time bin, respectively.
The epoch zero of each time bin is determined at the central point, and the spin frequency at such an epoch ($f(t_0)$) is identified through the $H$-test \citep{DB2010} and $Z_n^2$-test ($n< 4$; \citealt{Buc83}) to examine the pulsation in a small blind-searching region of ($f$, $\dot{f}$) inferred from the ephemerides provided by the previous studies
(\citealt{Ray2011,Lin&Hui} and the \emph{Fermi}\setcounter{footnote}{2}   team\footnote{http://www.slac.stanford.edu/$\sim$kerrm/fermi\_pulsar\_timing/J2021+4026/html/J2021+4026\_54683\_56587\_chol.par}).
 The uncertainties of the measured spin frequency and its derivative are expected to be smaller than their Fourier width, $\sim 8.3\times 10^{-8}$~Hz and $\sim 1.4\times 10^{-14}$~Hz s$^{-1}$, respectively.
The evolution of the frequency derivative ($\dot{f}$) obtained in A13 and this work is presented in the bottom panel of Fig.~\ref{evolution}, and the accuracy of the measured timing parameters can be improved by the time-of arrival (TOA) analysis \citep{Ray2011} of each individual time bin.

\begin{figure}[t]
  \centerline{\psfig{figure=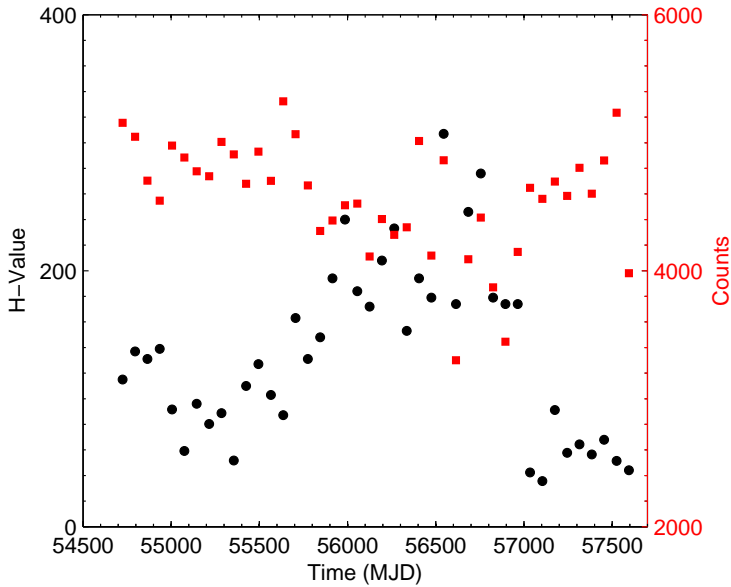, width=10cm,clip=}}
  \caption[\textcolor{}Evolution of significance of the pulsed detection for \psr ]{Evolution of corresponding $H$-statistics detected for the  pulsation of \psr. The black dots and red squares denote the detection significance in the  $H$-values and source counts in each 70-day time bin. A comparison of the detection significance between the pre-glitch/post-relaxation stages and the low-flux stage can clearly be seen as well.}
  \label{Hs}
\end{figure}

According to Fig.~\ref{evolution}, we see the coincident change in the gamma-ray flux and the frequency derivative of the spin.
Both sudden drops in the flux and the frequency derivative can be detected around $\sim$ MJD~55,850 and gradually returned to the original level passing the relaxation stage (i.e. $\sim$ MJD~56,950--57,050).
If we do not account for the frequency derivatives detected during the relaxation stage, the average of the  frequency derivatives assessed after the
relaxation stage is $-7.761(3)\times 10^{-13}$~Hz s$^{-1}$, which is close
to the pre-glitch value obtained in A13.We also investigate the evolution of the detection significance of the pulsation by comparing the $H$-values of each time bin. As shown in Fig. 3, we can obviously find that the pulsation of PSR J2021+4026 is much more significant in the low-flux stage.
During the low-flux stage (MJD 55,850-57000), the average number of
gamma-ray events collected to the target within 70 days is only 4240, and it has a significant pulsation with an $H$-value larger than 170. After the relaxation stage, the average number of photons extracted for PSR J2021+4026 are 4663 due to the increase of the gamma-ray flux, but most of
the corresponding $H-$values are between 35–-65.
Before the glitch, the average number of  gamma-ray photons collected
to the target within the same time interval is 4848, and the corresponding detection significance is still lower than that yielded in the low-flux stage.
It may provide an indication that the pulse fraction in the  low-flux stage
is higher than that in the pre-glitch and the post-relaxation stages.

\subsubsection{Ephemerides of \psr}
\label{sssec:ephemeris}

The \emph{Fermi} team provided a long-term ephemeris to cover the timing solution of \psr\ within MJD~54,670--56,580, including the flux jump around MJD~55,850 mentioned in A13.
To remove the effect of the timing noise, the pseudo-sinusoidal model is used to whiten the timing residuals \citep{Hobbs2004}.

{Because we want to compare the change of timing parameters at different epochs, we build two local ephemerides to describe the timing behavior of PSR J2021+4026 in the low-flux and post-relaxation stages, while the high-order derivatives of the pulse frequency are served as the parameters to account for the contribution of the timing noise.
At the beginning, we determine the spin parameters of \psr\ including only the spin-frequency and the spin-down rate (i.e., $f$ and $\dot{f}$) with a short time span of 100 days to avoid the influence of the accumulated timing noise.}
We build an initial template according to the Gaussian kernel density estimation (KDE) method \citep{DRS86} provided in \citet{Ray2011}.
By cross-correlating the template with the unbinned geocentered data, the pulse TOAs can be determined to fit to an original timing model assumed from the semi-blind search (``semi'' means that we just examined the pulsation of \psr\ in the vicinity of the known timing solution).
We polish the fit of the TOAs with TEMPO2 \citep{Hobbs2006,EHM2006}.
Once the timing residuals have  been minimized, the data can be extended to cover a longer time interval to generate more TOAs in modifying timing parameters.
If the fit has a significant aberration, we  include a phase jump or more high-order terms for the description of the timing noise in order to polish the fit.
\begin{table*}[t]
  \begin{center}
    \caption[]{Local ephemerides of \psr\ derived from LAT data of MJD~55,857--56,943 (Determined by 69 TOAs) and 57,062--57,565 (Determined by 47 TOAs). The numbers in parentheses denote 1$\sigma$ errors in the last digit.\\}\label{ephemerides}
    \begin{tabular}{lll}
      \hline\hline
      \multicolumn{2}{l}{Parameter} \\
      \hline
      Right ascension, $\alpha$\dotfill & \multicolumn{2}{c}{20:21:30.733} \\
      Declination, $\delta$\dotfill &  \multicolumn{2}{c}{+40:26:46.04} \\
      Valid MJD range\dotfill & 55,857--56,943 & 57,062--57,565 \\
      Pulse frequency, $f$ (s$^{-1}$)\dotfill & 3.7689669240(2) & 3.7689112482(6) \\
      First derivative of pulse frequency, $\dot{f}$ (s$^{-2}$)\dotfill & $-8.1978(1)\times10^{-13}$ & $-7.738(1)\times10^{-13}$ \\
      second derivative of pulse frequency, $\ddot{f}$ (s$^{-3}$)\dotfill & $-1.9(2)\times10^{-22}$ & $3.5(3)\times10^{-22}$ \\
      Third derivative of pulse frequency, $\dddot{f}$ (s$^{-4}$)\dotfill & $5.72(7)\times10^{-29}$ & $-1.7(9)\times10^{-29}$\\
      Fourth derivative of pulse frequency, $f^{(4)}$ (s$^{-4}$)\dotfill & $3(1)\times10^{-38}$ & $-6(9)\times10^{-37}$\\
      Epoch zero of the timing solution (MJD)\dotfill & 56,400 & 57,200\\
      RMS timing residual ($\mu$s)\dotfill & 2199.970 & 2795.366\\
      Time system \dotfill & \multicolumn{2}{c}{TDB (DE405)} \\
      \hline
    \end{tabular}
  \end{center}
  \end{table*}

\begin{table*}
  \centering
  \begin{tabular}{ccccc}
    \hline\hline
    & Pre-glitch & Low-flux & \multicolumn{2}{c}{Post-relaxation} \\
    \hline
    Gaussian components& Three\tablenotemark{c} & Two\tablenotemark{c}  & Two\tablenotemark{d}  & Three\tablenotemark{d}  \\
    Peak 1\tablenotemark{a}\dotfill & $0.19\pm 0.02$ & $0.13\pm 0.02$ & $0.195\pm 0.024$ & $0.205\pm 0.024$ \\
    Peak 2\tablenotemark{a}\dotfill & $0.176\pm 0.007$ & $0.174\pm 0.006$ & $0.152\pm 0.010$ & $0.16\pm 0.009$ \\
    Peak 3\tablenotemark{a}\dotfill & $0.11\pm 0.02$ & - & - & $0.073\pm 0.045$ \\
    
    Peak 1/Peak 2\tablenotemark{b}\dotfill & $0.54\pm 0.06$ & $0.24\pm 0.03$ & $0.494\pm 0.009$ & $0.496\pm 0.008$ \\
    Peak 3/Peak 2\tablenotemark{b}\dotfill & $0.16\pm 0.03$ & - & - & $0.12\pm 0.06$ \\
    &  &  &  &  \\
    $\chi^2$/D.O.F. &  &  & 32.1/43 & 29.1/40 \\
    \hline
        {\footnotesize a: FWHM~~~~~~~~~~~~~~~~~~} &&&&\\
        {\footnotesize b: Ratio of amplitude~~~} &&&&\\
        {\footnotesize c: Allafort et al. (2013)} &&&&\\
        {\footnotesize d: This work~~~~~~~~~~~~~~} &&&&\\
  \end{tabular}
  \caption{Parameters of Gaussian fitting for pulse profiles of $E >0.1$~GeV at the pre-glitch and the
    low-flux stages, and the post-relaxation (this work).
    Two and three Gaussian components fitting are represented for  the
    post-relaxation.}
\end{table*}

With such an iteration process, we finally generate two solutions including $f, \dot{f}, \ddot{f}, \dddot{f}$, and $f^{(4)}$ to describe the timing behavior of \psr\ during $\sim$MJD~55,850--56,950 (within the low-flux stage) and 57,050--57,570 (after the relaxation stage).
The results are summarized in Table~\ref{ephemerides}.
We also note that these solutions cannot well describe the timing behavior during the relaxation stage (i.e. $\sim$MJD~56,950--57,050).
If we would like to slightly extend the local timing solution to cover the data epoch in relaxation, at least two more high-order polynomial terms are required to obtain an acceptable fit of the determined TOAs.
This means that the data within the epoch of the relaxation stage (i.e. $\sim$MJD~56,950--57,050) have an effect due to a real high-order timing term,
a glitch,  or serious short-term timing noises.
The identification of the time range of the relaxation is consistent with the required time for \psr\ to bring its gamma-ray flux back to the original level, as shown in Fig~\ref{evolution}.

\subsection{Pulse Profiles and Phase-Resolved Spectra}
\label{pulsef}
\begin{figure}
  \centering
  \epsscale{1}
\plotone{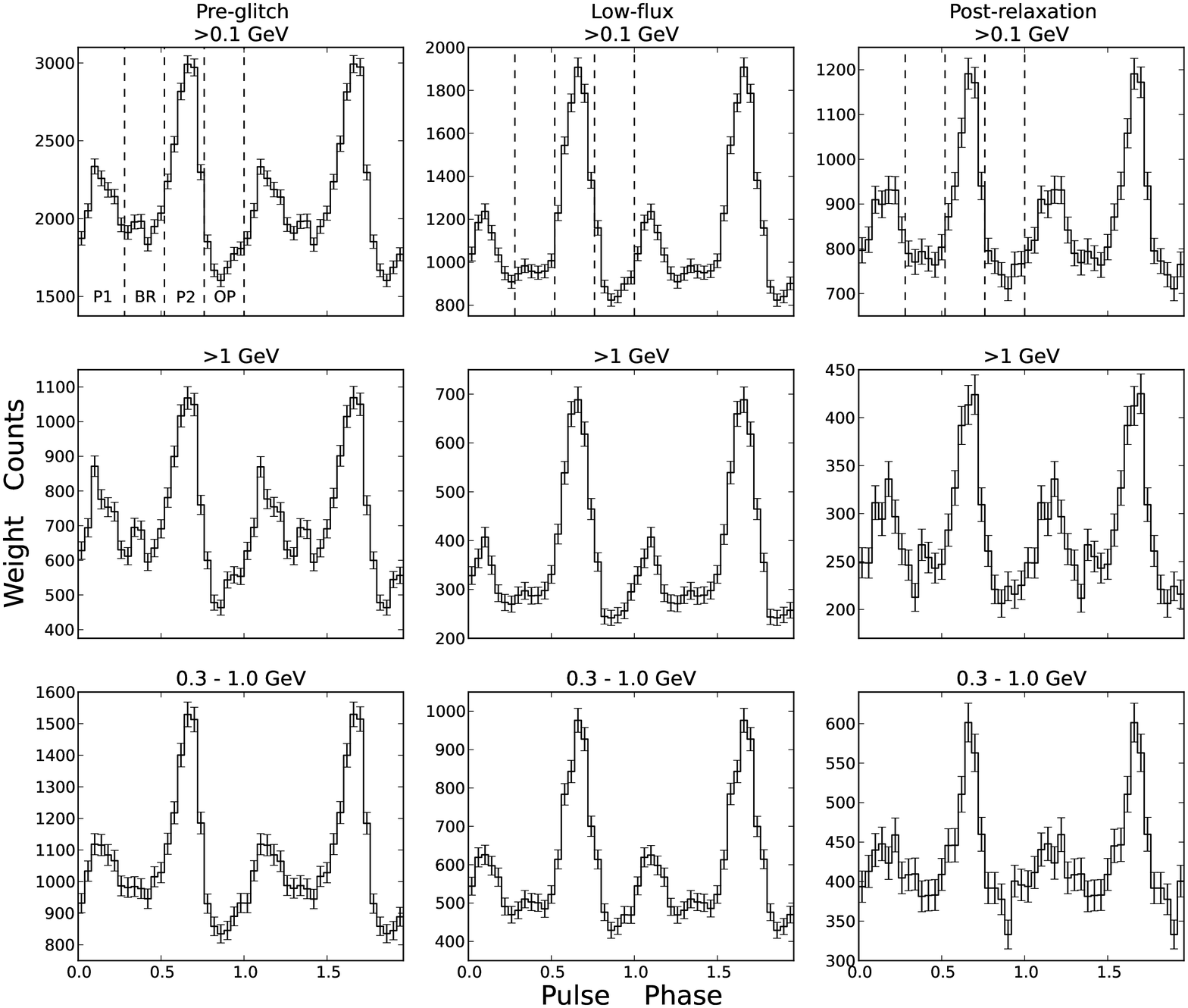}
\caption{Pulse profiles of PSR~J2021+4026. From left to right are the  pulse
  profiles for the pre-glitch, low-flux, and post-relaxation stages.
  From top to bottom are the pulse profiles generated with photon energy $>0.1$~GeV,  $>1$~GeV, and $0.3-1$~GeV.   The photons used  are collected in  MJD54,589-55,850 for the pre-glitch stage,  MJD55,850-56,580
    for the low-flux stage, and MJD57,062-57,565 for the post-relaxation stage.
  The total ``weighted'' count and total count are, respectively, 
  $N_{>0.1GeV}\sim (5.3\times 10^4, 8.0\times 10^4)$, $N_{>1GeV}\sim (1.8\times 10^4, 2.0\times 10^4, )$ and $N_{0.3-1GeV}\sim(2.7\times 10^4, 3.5\times 10^4$) for the pre-glitch stage;
  $N_{>0.1GeV}\sim (2.8\times 10^4, 4.5\times 10^4)$, $N_{>1GeV}\sim (0.9\times 10^4, 1.1\times 10^4)$
  and $N_{0.3-1GeV}\sim (1.5\times 10^4 , 2.0\times 10^4)$ for the  low-flux stage, and
  $N_{>0.1GeV}\sim (2.2\times 10^4, 3.2\times 10^4)$,   $N_{>1GeV}\sim (0.7\times 10^4,
  0.8\times 10^4)$ and $N_{0.3-1GeV}\sim (1.1\times 10^4, 1.4\times 10^4)$
  for the post-relaxation stage. The vertical dashed lines in the top panels indicate the phases of the first peak (P1), bridge region (BR),
  second peak (P2) and off-pulse region (OP); (P1, BR, P2, OP)=(0-0.28, 0.28-0.52 0.52-0.76,0.76-1). }
  \label{pulse}
 \end{figure}

Using the timing solution provided by the \emph{Fermi} team and the local ephemerides summarized in Table~\ref{ephemerides}, we generate the pulse profiles
with the \emph{Fermi} archive for the pre-glitch stage, low-flux stage, and
post-relaxation stage.  We select photons within a 1$^{\circ}$ radius of the pulsar and produce  pulse profiles using the best-fit spectrum to assign each photon a weight, which denotes the probability of a photon originating from the pulsar. 
We assigned the phase of each photon with timing solution provided by \emph{Fermi} team for the pre-glitch and the low-flux stages and by Table~\ref{ephemerides} after the relaxation.
The phase of each pulse profile is shifted to align the major/second peak for three different stages.
Fig.~\ref{pulse} shows the energy-dependent pulse profiles in three stages.

A13 found that the $>0.1$GeV pulse profile shows a third peak in the bridge of the pre-glitch stage,
and they described the profile with three Gaussian components. By contrast,
two Gaussian fits were sufficient to describe the pulse profile of the low-flux stage (see Table~3).
To compare with their results, we fit the pulse profile after the relaxation
with two and three Gaussian components, and we present the parameters in Table~3. The null hypothesis probability in the $F$-test to support the requirement of the third Gaussian component is less than 1$\sigma$, and we cannot conclude the recovery of a third component/peak at the BR after the
relaxation. By comparing the fitting parameters at the pre-glitch and
post-relaxation stages, however, we notice that the FWHMs (Full Width at Half Maximum) and the peak ratios of the three Gaussian components of the post-relaxation stage
are consistent with those of the pre-glitch stage within a  1$\sigma$ level. We also note that the energy-dependent light
curves of the post-relaxation stage are  more consistent with those of the pre-glitch stage. 

To determine the phase-resolved spectra for the  three stages, we divide the profiles  into four phase intervals: the first peak (P1), the BR, the second peak (P2) and the off-pulse region (OP) are defined by (P1, BR, P2, OP)=(0-0.28, 0.28-0.52, 0.52-0.76,0.76-1).  
For each phase, we determine the spectrum over 10 logarithmically distributed energy bins from 100~MeV to 100~GeV.
If the detection significance is less than 3$\sigma$, an upper limit is given to describe the spectral behavior.
We fit each spectrum with a power law plus an exponential cutoff function.

\begin{figure}
  \centering
  \plotone{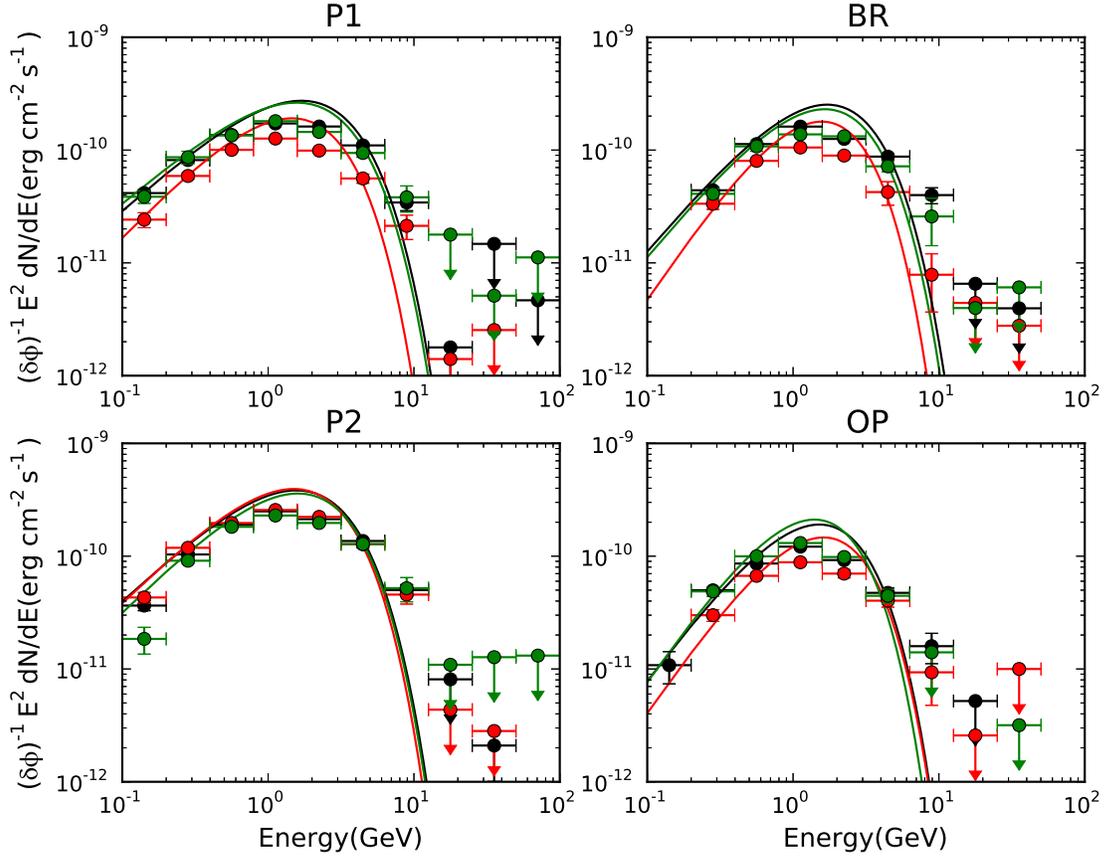} 
  \caption{Phase-resolved spectra at the three different stages. The black data points and curves present the spectral behavior in the pre-glitch stage, while the red and green ones denote the spectra of the low-flux and post-relaxation stages, respectively.}
  \label{phase}
\end{figure}

The Fig.~\ref{phase} shows the phase-resolved spectra. We can see in Fig.~\ref{phase} that
the spectra of the post-relaxation stage (black) are more consistent with those of the pre-glitch stage (green), indicating that  the structure of the particle-accelerating region was also recovered after relaxation. We also note
that the variability of the second/major peak is significantly smaller than that of the other phases among the three stages.

\section{Summary and Discussion}
\begin{figure}
  \centering
  \epsscale{1.0}
  \plotone{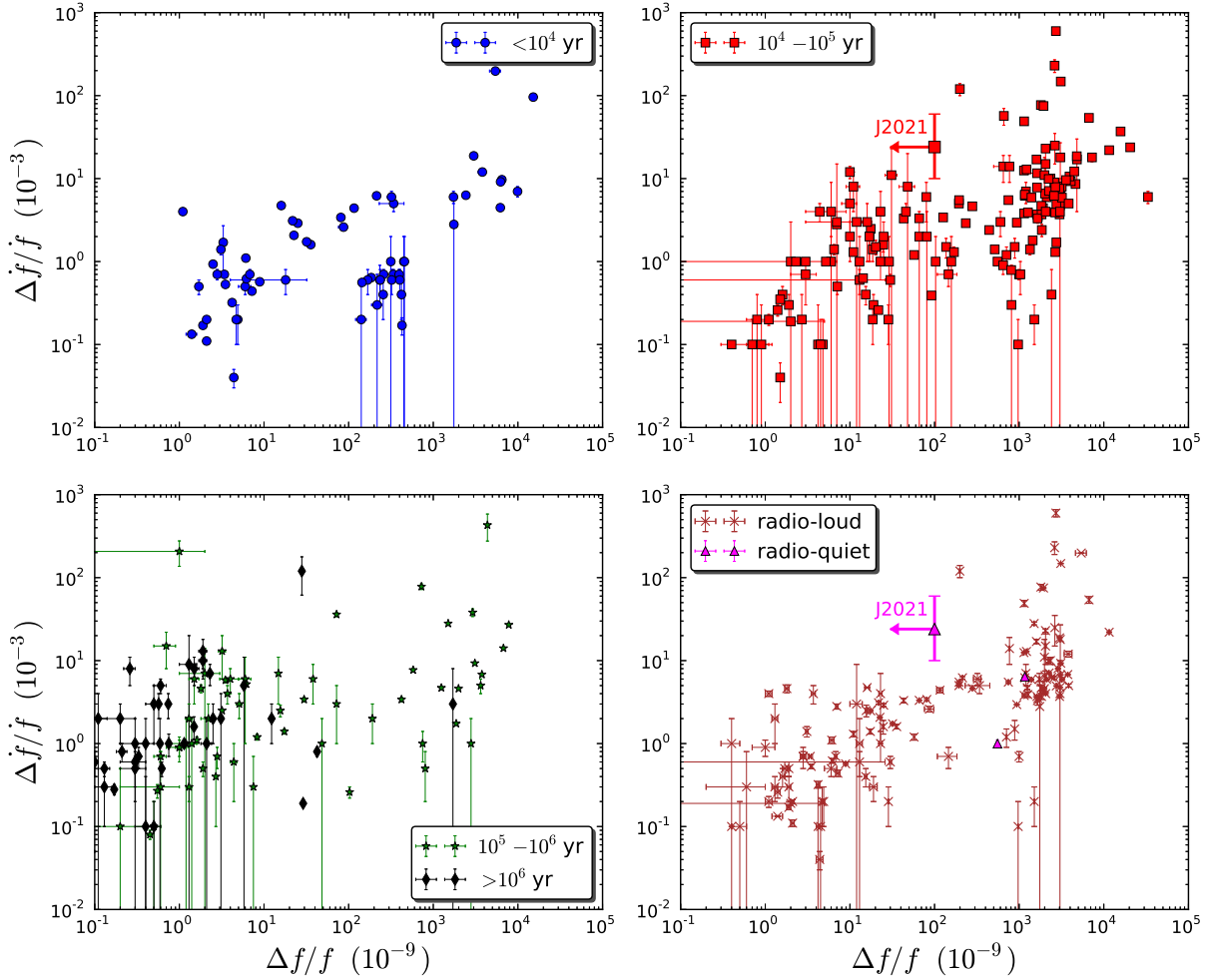}
  \caption{Fractional glitch size of the known pulsars. The data are taken from the glitch catalog (Espinoza et al. 2011; http://www.jb.man.ac.uk/pulsar/glitches.html). The different panels (top panels and bottom left panel)
    represent the different characteristic ages of the pulsars (Manchester et al. 2005).
    The glitch of PSR~J2021+4026 (a larger symbol  with an upper  limit of the glitch size) is also indicated in this figure. The bottom right  panel shows the glitch size for the   gamma-ray pulsars. Of the radio-quiet pulsars, 
      only three (J0007+7303, J1813-1246, and J2021+4026) are presented, since $\triangle \dot f$  for  other sources has not been measured or cannot be presented in the figure (Table~4).}
  \label{size}
\end{figure}
\label{disc}

\begin{table}
\begin{tabular}{cccccccc}
  \hline\hline
  PSRs & MJD & $\triangle f/f$ & $\triangle \dot{f}/\dot{f}$ & $P$ & $B_s$ &$L_{sd}$ & $\
  \tau$ \\
  J &  & $10^{-9}$ & $10^-3$ & s & $10^{12}$G &$10^{35}$erg/s & $10^4$yr \\
  \hline
  0007+7303 & 54953 & 554 & $1\pm 0.1$ & 0.32 &11 & 4.5 & 1.4 \\
  0007+7303 & 55466 & 1260 & - &  0.32 &11 &4.5 & 1.4 \\
  1023-5746 & 55041 & 3560 & - & 0.11 &6.6 & 110 & 0.46 \\
  1413-6205 & 54735 & 1730 & - &0.11  &1.8 & 8.3 & 6.3 \\
  1813-1246 & 55094 & 1166 & 6.4 & 0.048 & 0.9 &62  & 4.3 \\
  1838-0537 & 55100 & 5500 & -15 & 0.15 &  8.4&60 & 0.49 \\
  1907+0602 & 55422 & 4660 & - & 0.11 &3.1 & 28 & 2.0 \\
  1907+0602 & 55866 & 1.49 & -0.26 & 0.11 &3.1 & 28  & 2.0 \\
  2021+4026 & $\sim$ 55850 & $<10^{2}$ &24 & 0.27& 3.9&1.2  & 7.7 \\
  2032+4127 & 55821 & 276.2 & 0$\pm 1$ & 0.14 &1.4& 1.7 & 18 \\
  \hline
\end{tabular}
\caption{Parameters of the glitch events and spin-down parameters of the radio-quiet gamma-ray
pulsars. The second column shows the day of the glitch events.}
\end{table}
We report the investigation of timing and spectral analysis with the \emph{Fermi}-LAT data of PSR J2021+4026 after the glitch occurred around 2011 October 16. We show that the pulsar stayed at a high spin-down
rate  and a low gamma-ray
flux state for about 3-yr after the glitch.
Around 2014 December, the spin-down rate and all properties of the gamma-ray emission returned to the pre-glitch state within a time scale of a few months. Our spectral and timing analysis suggests that
the pulsar magnetosphere suffered a state change triggered  by the glitch in 2011 and returned to the pre-glitch state around 2014 December. 

A glitch appears as an abrupt increase in the spin frequency, $f$, and in the spin-down rate, $\dot{f}$, and it is usually observed with 
 a  size of  $\Delta f/f\sim 10^{-10}-10^{-4}$ and  $|\Delta \dot{f}/\dot{f}|\sim 10^{-4}-1$, respectively \citep{esp11}. 
More than 160 glitching pulsars have been confirmed\footnote{http://www.jb.man.ac.uk/pulsar/glitches.html}, and about 50 \emph{Fermi}-LAT pulsars have shown glitch events.  The observed fractional glitch size $\Delta f/f$ shows a bimodal distribution \citep{wang10, esp11, yu13} for which the first peak and
second peak appear at $\Delta f/f\sim 10^{-9}$ and $\sim 10^{-6}$ (see Fig.~\ref{size}), respectively, suggesting two kinds of glitch mechanisms.
It has been considered that the glitch of a pulsar is attributed to the release of stress buildup as a result of the steady spin-down.  This stress is stored in the solid crust of a star \citep{rud69, bay71, has15} and/or on pinned vortices in the superfluid interior \citep{alp84, has15}.
The crust rearrangement by the starquakes could induce the small glitches.
However, the glitches of the Vela pulsar have been observed every 2 or 3 yr
with a size of $\Delta f/f\sim 10^{-6}$, which is too huge to be explained by the stress stored at recurrence time.
It has been suggested that a sudden momentum transfer from a faster rotation interior superfluid to the solid crust is the origin of the observed large glitches.

For PSR~J2021+4026, the apparent change in the spin frequency was not detected by the \emph{Fermi}-LAT (see A13), but we may read the glitch size as $\Delta f/f < 10^{-7}$ from Fig.~2 and~3 in A13.  
We can see in Fig.~\ref{size} that old pulsars with a spin-down age $\tau>10^6$~yr mainly show a small glitch size with $\Delta f/f<10^{-8}$, while younger pulsars with $\tau <10^{6}$ yr exhibit the glitch sizes in a wide range of $10^{-10}<\Delta f/f<10^{-4}$. With $\tau\sim 77$~kyr of PSR~J2021+4026, therefore, we cannot constrain the glitch mechanism from the $\Delta f/f-\tau$ relation in Fig.~\ref{size}.   
The figure also shows that a relatively large jump in the spin-down rate ($\Delta \dot {f}/\dot{f}>10^{-2}$) is mainly  accompanied by a large glitch size ($\Delta f/f>10^{-6}$), indicating the origin of the angular momentum transfer from the superfluid to the solid crust. 
The glitch of PSR~J2021+4026 was observed with a jump, $\Delta \dot {f}/\dot{f}\sim 0.01-0.06$ (see A13), which is relatively large compared to glitches of the other pulsars with $\Delta f/f < 10^{-7}$. 

  The bottom right panel of Figure~\ref{size} shows the glitch sizes of the $Fermi$-LAT gamma-ray pulsars; the crosses  and triangles  represent radio-loud and radio-quiet gamma-ray pulsars, respectively.
  Of the radio-quiet pulsars, only three (J0007+7303, J1813-1246 and J2021+4026) are presented,
  since $\triangle \dot{f}$  for other sources has not been measured or cannot be presented in the figure (see Table 4).
  As Table~4 indicates, PSR~J2021+4026 has the  largest jump in the spin-down rate at the glitch,  the smallest
  spin-down power, and a larger spin-down age compared to other  glitching  radio-quiet pulsars. By contrast, among all radio-quiet
  gamma-ray pulsars,  PSR~J2021+4026 has typical spin-down parameters.
  It may be worth noting that the observed gamma-ray-to-X-ray flux ratio $F_{\gamma}/F_X\sim 6.5\times 10^4$
  of PSR~J2021+4026   is the largest among the $Fermi$-LAT pulsars (Hui et al. 2017)

One of the intriguing properties of \psr\ is that the gamma-ray flux drops at the glitch.  A few pulsars exhibit a radiative event at the glitch. 
PSR~J1119-6127 is the high-magnetic-field radio pulsar, and
four glitches of this pulsar have been observed.
The pulsar showed an additional component in the radio emission following a large-amplitude glitch \citep{wel11, ant15}, as well as a magnetar-like X-ray burst in 2016 July, which was probably
associated with a glitch \citep{arc16, gog16}.
Its pulsed radio emission disappeared after the burst and reappeared
about two weeks after the event \citep{bur16, maj17}. The pulsed gamma-ray emission also might be affected by this event \citep{tam16}. 
PSR~J0742-2822 shows the radio pulse profile varying on a time-scale of decades \citep{lyn10, kar11}, and it's mode-switching is probably linked to glitch events (Keith et al. 2013) as well. These observations suggest  some glitches  affect the particle acceleration and emission processes in the pulsar magnetosphere.  

Although the pulsars show a variety of the post-glitch recoveries \citep{yua10,esp11, yu13}, the post-glitch behavior of PSR~J2021+4026 is unique.
Most of the glitches show a relaxation toward the timing solution extrapolated from the pre-glitch
behaviors.
Exponential changes in both $f$ and $\dot{f}$ with time are often observed immediately after the glitch, and its time scale is $\sim 100$~days.
The post-glitch relaxation of the Crab pulsar is  described well  by this exponential recovery. 
In a longer-term recovery, the post-glitch behavior is dominated by a linear increase of the spin-down rate $\dot{f}$, and this linear recovery sometimes continues until the next glitch.
Because of the sensitivity of the \emph{Fermi}-LAT, it is difficult to discuss whether PSR~J2021+4026 showed an exponential recovery immediately after the glitch.
In the long-term behavior, by contrast, the current result (Fig.~\ref{evolution}) shows that the pulsar stayed at a high spin-down rate and a low gamma-ray flux state for about 3 yr after the glitch. 
This implies that the glitch event in 2011 October triggered a mode change in the spin-down rate and gamma-ray emission.  
It has been considered that the gamma-ray emission from the pulsar is produced in an acceleration region near the light cylinder \citep{abdo09c};hence the observed coincident evolution in the spin-down rate and gamma-ray flux of PSR~J2021+4026 (Fig.~\ref{evolution}) suggests that the glitch event in 2011 October affected the magnetospheric structure around the light cylinder.

Several pulsars show  that the glitch triggers a permanent change in the spin-down rate ($\dot{f}$); for examples, the Crab pulsar \citep{lyn13, lyn15} and
PSRs B0144+59 and B0402+61 \citep{yua10, yu13}.  
The permanent shift in spin-down rate may be  a result of a  change in
the magnetic axis orientation,   magnetospheric current flow and/or  moment of inertia due to rearrangement of the crust \citep{rud98, ant15}.
For PSR~J2021+4026, a change in the moment of inertia will be limited by  angular momentum conservation at the glitch $|\Delta I|/I=|\Delta f|/f< 10^{-7}$, which will not affect the observed permanent change in the spin-down rate $\Delta \dot{f}/\dot{f}>0.01$.
With the gamma-ray flux drop, and therefore, the glitch of PSR~J2021+4026 likely triggered the mode change of the global magnetosphere.

The glitch of the Crab pulsar has been observed with a permanent shift in the spin-down rate, and each glitch increases by an order of $|\Delta \dot{f}|\sim 10^{-13}$ \citep{lyn15}.
This could be interpreted as a consequence of the increase in the angle between the rotation axis and the magnetic axis.  The low breaking index $n=2.35$ and the increase in the phase separation between the main pulse  and interpulses
 of the Crab pulsar might be due to the increase in the magnetic inclination angle \citep{lyn13, lyn15}.

\citet{ng16} stated that the permanent-like change in the spin-down rate of PSR~J2021+4026 is due to the increase of the magnetic inclination angle. 
The spin-down luminosity of the force-free magnetosphere was shown by \citet{spi06}, and it is related to the magnetic inclination angle ($\alpha$) as
\begin{equation}
  L_{sd}\sim \frac{\mu^2\Omega^4}{c^3}(1+\sin^2\alpha), 
\label{lsd}
\end{equation}
where $\mu$ is the dipole momentum and $\Omega=2\pi f$. Replacing with  $L_{sd}=I\Omega\dot{\Omega}$, where $I$ is the  moment of inertia of the neutron star, we find that  the relative size in the spin-down rate can be expressed as
\begin{equation}
  \frac{\Delta \dot{f}}{\dot{f}}=\frac{\sin 2\alpha \Delta \alpha}{1+\sin^2\alpha}.
\end{equation}
Assuming $\alpha\sim 63^{\circ}$ before the glitch of \psr, which was assumed in \citet{ng16} to fit the observed spectra and pulse profiles, the inclination angle was increased by $\sim 5^{\circ}$. 

Alternatively, we speculate that the glitch of PSR~J2021+4026 could change the magnetic field structure in the polar cap region and subsequently cause a change in the magnetospheric current flow. 
The increase in the spin-down torque after the glitch may correspond to an increase in the current flow by an order of $\delta i/i\sim \Delta \dot{f}/\dot{f}\sim 0.01$, where $i$ is the total current circulating in  the
magnetosphere.  It has been discussed that the pair-creation process of the high-energy gamma rays maintains the global current circulating in the pulsar magnetosphere \citep{wad11,yuk12}.
In the polar cap acceleration region, the current carriers are supplied by the magnetic pair-creation process and photon-photon pair-creation process \citep{dau96}. The mean-free path of the magnetic pair creation is expressed by  
\[ 
  \ell=\frac{4.4}{(e^2/\hbar c)}\frac{\hbar}{mc}\frac{B_Q}{B\sin\theta}
  {\rm exp}\left(\frac{4}{3\chi}\right)
      \]
\citep{rud75, tim15} with
\[
\chi=\frac{E_{\gamma}}{2m_ec^2}\frac{B\sin\theta}{B_Q}
\]
where $B$ is the local magnetic field strength, $B_Q=m_e^2c^2/e\hbar=4.4\times 10^{13}$G, $E_{\gamma}$ is the energy of the photon, and $\theta$ is the angle between the propagation direction for the photon and the magnetic field. 
The mean-free path of the process depends exponentially on the field strength and propagation direction.
If the glitch event of PSR~J2021+4026 triggered the change in the magnetic field structure (dipole and/or multipole fields) of the polar cap region, the global current flow and hence the spin-down torque might be affected.
Accordingly, such a change in the current flow can induce a stage change in the gamma-ray emission \citep{tak16}. 

Our phase-resolved  analysis (Fig.~\ref{phase}) shows that
the variability of the second peak (P2) is significantly smaller
than that of the other pulse phases among the three stages.
Ng et al. (2016) demonstrated this behavior with the outer gap model. 
Within the framework of the outer gap model, this could be explained by
the effect of three-dimensional geometry.  The outer gap  model expects that the particle accelerator
and emission region  extend from the null charge surface of the Goldreich-Julian charge density to
the light cylinder along the magnetic field lines; the Goldreich-Julian charge density is defined by -$\mathbf{\Omega}\cdot \mathbf{B}/2\pi c$ \citep{gol69} and the null charge surface corresponds to $\Omega\cdot B=0$. The three-dimensional
studies of the outer gap (Romani and   Yadigalogru 1995; Cheng et al. 2000; Tang et al. 2008)
have predicted that only emission around the light cylinder (radial distance
$r\sim R_{lc}=c/\Omega$) makes  the first peak, while  the
emission from various   regions, $r\sim (0.1-1)R_{lc}$, contributes to the second peak. 
 Since the emission from a narrower range of  radial distances
 contributes to the first peak and BR (maybe off-pulse emission), the change of the outer gap structure
  triggered by the glitch   directly  affects  the observed emission.  For the second peak, by contrast,  
the contributions of the emission from the various radial distances may reduce the effect of the state change of the
outer gap on the observed spectrum.

Switching between discrete states of the magnetosphere may be common in  pulsar populations. 
Some pulsars have shown abrupt changes between two states of the radio emission, so-called ``mode-changing'' or ``null'' \citep{wan07}, and PSR~B0943+10
shows synchronous mode-switching between the radio and the X-ray emission \citep{her13, mer16}. 
Two emitting states for some pulsars have been observed with different spin-down states \citep{kra06,lyn10}.
For example, PSR B1931+24 is switching between radio-on and radio-off states lasting for days to weeks, in which the spin-down rate of the ``on state'' is
a factor of 1.5 larger than that of the ``off state''. 
The mode change in the radio emission has been interpreted as the state change of the magnetospheric current \citep{kra06}, although its trigger has not been understood yet.
PSR~J2021+4026 may serve as  another example of  a mode change of the radiation due to the state change of the global magnetosphere, and a long-term monitoring of the timing parameter of this target may provide us a hint to understand the process of the mode change. 

 We thank the referee for the comments that improved this paper.
 JZ and  JT are supported by NSFC grants of Chinese Government under 11573010,
 U1631103 and 11661161010. 
CWN and KSC are supported by GRF grant under 17302315. PHT is supported by the National Science Foundation of China (NSFC) through grants 11633007 and 11661161010. CYH is supported by the National Research Foundation of Korea through grants 2014R1A1A2058590 and 2016R1A5A1013277. AKHK is supported by the Ministry of Science and Technology of Taiwan through grant 105-2112-M-007-033-MY2, 105- 2119-M-007-028-MY3, and 106-2918-I-007-005. DCCY is supported by the Ministry of Science and Technology of Taiwan through grant 105-2115-M-030-005-MY2


\begin{thebibliography}{}
  \expandafter\ifx\csname natexlab\endcsname\relax\def\natexlab#1{#1}\fi
\bibitem[{Abdo} {et~al.}(2009a)]{abdo09a}
  {Abdo}, A.~A., {Ackermann}, M., {Ajello}, M., {Atwood}, W.~B., {Axelsson}, M. {et~al.},
  2009a, \apjs, 183, 46

\bibitem[{Abdo} {et~al.}(2009b)]{abdo09b}
  {Abdo}, A.~A., {Ackermann}, M.,  {Ajello}, M.,  {Anderson}, B.,
  {Atwood}, W.~B., {et~al.}, 2009b, Science, 325, 840

\bibitem[{Abdo} {et~al.}(2009c)]{abdo09c}
  {Abdo}, A.~A., {Ackermann}, M., {Atwood}, W.~B. , {Bagagli}, R. ,
  {Baldini}, L., {Ballet}, J., {Band}, D.~L., {Barbiellini}, G.,
  {Baring}, M.~G., {Bartelt}, J., et al., 2009c, ApJ, 696, 1084
\bibitem[{Acero} {et~al.}(2015)]{acero15}
  {Acero}, F., {Ackermann}, M., {Ajello}, M., {Albert}, A.,
  {Atwood}, W.~B. and {Axelsson}, M. {et~al.}, 2015, \apjs, 218, 23

\bibitem[{{Allafort} {et~al.}(2013){Allafort}, {Baldini}, {Ballet},
    {Barbiellini}, {Baring}, {Bastieri}, {Bellazzini}, {Bonamente}, {Bottacini},
    {Brandt}, {Bregeon}, {Bruel}, {Buehler}, {Buson}, {Caliandro}, {Cameron},
    {Caraveo}, {Cecchi}, {Chaves}, {Chekhtman}, {Chiang}, {Chiaro}, {Ciprini},
    {Claus}, {D'Ammando}, {de Palma}, {Digel}, {Di Venere}, {Drell}, {Favuzzi},
    {Ferrara}, {Franckowiak}, {Fusco}, {Gargano}, {Gasparrini}, {Giglietto},
    {Giroletti}, {Glanzman}, {Godfrey}, {Grenier}, {Guiriec}, {Hadasch},
    {Harding}, {Hayashida}, {Hayashi}, {Hays}, {Hewitt}, {Hill}, {Horan}, {Hou},
    {Jogler}, {Johnson}, {Johnson}, {Kerr}, {Kn{\"o}dlseder}, {Kuss}, {Lande},
    {Larsson}, {Latronico}, {Lemoine-Goumard}, {Longo}, {Loparco}, {Lubrano},
    {Malyshev}, {Marelli}, {Mayer}, {Mazziotta}, {Mehault}, {Mizuno}, {Monzani},
    {Morselli}, {Murgia}, {Nemmen}, {Nuss}, {Ohsugi}, {Omodei}, {Orienti},
    {Orlando}, {Paneque}, {Pesce-Rollins}, {Pierbattista}, {Piron}, {Pivato},
    {Porter}, {Rain{\`o}}, {Rando}, {Ray}, {Razzano}, {Reimer}, {Reposeur},
    {Romani}, {Sartori}, {Saz Parkinson}, {Sgr{\`o}}, {Siskind}, {Smith},
    {Spinelli}, {Strong}, {Takahashi}, {Thayer}, {Thompson}, {Tibaldo},
    {Tinivella}, {Torres}, {Tosti}, {Uchiyama}, {Usher}, {Vandenbroucke},
    {Vasileiou}, {Venter}, {Vianello}, {Vitale}, {Winer}, \&
    {Wood}}]{Allafort2013}
  {Allafort}, A., {et~al.} 2013, \apj Letter, 777, L2
\bibitem[{Alpar} {et~al.}(1984)]{alp84}
  {Alpar}, M.~A., {Pines}, D., {Anderson}, P.~W. and {Shaham}, J., 1984, ApJ, 276, 325
\bibitem[{Antonopoulou} {et~al.}(2015)]{ant15}
  {Antonopoulou}, D., {Weltevrede}, P.. {Espinoza}, C.~M.,
  {Watts}, A.~L.,  {Johnston}, S., et al., 21015, \mnras, 447, 392
\bibitem[{Archibald} {et~al.}(2016)]{arc16}
  {Archibald}, R.~F., {Kaspi}, V.~M., {Tendulkar}, S.~P. and
  {Scholz}, P., 2016, \apj Letter, 829, 21
\bibitem[{Baym and Pines}(1971)]{bay71}
  {Baym}, G. and {Pines}, D., 1971, Annals of Physics, 66, 816

\bibitem[{{Buccheri} {et~al.}(1983){Buccheri}, {Bennett}, {Bignami}, {Bloemen},
    {Boriakoff}, {Caraveo}, {Hermsen}, {Kanbach}, {Manchester}, {Masnou},
    {Mayer-Hasselwander}, {Ozel}, {Paul}, {Sacco}, {Scarsi}, \& {Strong}}]{Buc83}
  {Buccheri}, R., {et~al.} 1983, \aap, 128, 245
\bibitem[{Burgay} {et~al.}(2016)]{bur16}
  {Burgay}, M., {Possenti}, A., {Sanna}, A., {Papitto}, A., {Burderi},
  L., 2016, The Astronomer's Telegram, 90740

\bibitem[{Cheng} {et~al.}(2000)]{ch00}
  {Cheng}, K.~S. and {Ruderman}, M. and {Zhang}, L., 2000, ApJ, 537, 964
\bibitem[{Daugherty and Harding}(1996)]{dau96}
  {Daugherty}, J.~K. and {Harding}, A.~K., 1996, \apj, 458, 278

\bibitem[{{de Jager} \& {B{\"u}sching}(2010)}]{DB2010}
  {de Jager}, O.~C., \& {B{\"u}sching}, I. 2010, \aap, 517, L9

\bibitem[{{de Jager} {et~al.}(1986){de Jager}, {Raubenheimer}, \&
    {Swanepoel}}]{DRS86}
  {de Jager}, O.~C., {Raubenheimer}, B.~C., \& {Swanepoel}, J.~W.~H. 1986, \aap,
  170, 187

\bibitem[{{Edwards} {et~al.}(2006){Edwards}, {Hobbs}, \&
    {Manchester}}]{EHM2006}
  {Edwards}, R.~T., {Hobbs}, G.~B., \& {Manchester}, R.~N. 2006, \mnras, 372,
  1549
\bibitem[{Espinoza} {et~al.}(2011)]{esp11}
  {Espinoza}, C.~M., {Lyne}, A.~G., {Stappers}, B.~W. and
  {Kramer}, M., 2011, \mnras, 414, 1689
\bibitem[{Goldreich and Julian} (1969)]{gol69}
  {Goldreich}, P. and {Julian}, W.~H., 1969, ApJ, 157, 869

\bibitem[{G{\"o}{\u g}{\"u}{\c s}} {et~al.}(2016)]{gog16}
  G{\"o}{\u g}{\"u}{\c s}, E., {Lin}, L., {Kaneko}, Y.,
  {Kouveliotou}, C.,  {Watts}, A.~L., et al.,  2016, \apj Letter, 829, 25
\bibitem[{Haskell and Melatos}(2015)]{has15}
  {Haskell}, B. and {Melatos}, A., 2015,  International Journal of Modern
  Physics D, 24, 15300
\bibitem[{Hermsen} {et~al.}(2013)]{her13}
  {Hermsen}, W., {Hessels}, J.~W.~T., {Kuiper}, L., {van Leeuwen}, J.,
  {Mitra}, D., et al., 2013, Science, 339, 436
\bibitem[{{Hobbs} {et~al.}(2004){Hobbs}, {Lyne}, {Kramer}, {Martin}, \&
    {Jordan}}]{Hobbs2004}
  {Hobbs}, G., {Lyne}, A.~G., {Kramer}, M., {Martin}, C.~E., \& {Jordan}, C.
  2004, \mnras, 353, 1311
\bibitem[{{Hobbs} {et~al.}(2006){Hobbs}, {Edwards}, \&
    {Manchester}}]{Hobbs2006}
  {Hobbs}, G.~B., {Edwards}, R.~T., \& {Manchester}, R.~N. 2006, \mnras, 369, 655
  bibitem[{Hobbs} {et~al.}(2017)]{hui17}
  {Hui}, C.~Y. and {Lee}, J. and {Takata}, J. and {Ng}, C.~W. and
  {Cheng}, K.~S., 2017, \apj, 834, 120
  
\bibitem[{Karastergiou}{et~al.}(2011)]{kar11}
  {Karastergiou}, A., {Roberts}, S.~J., {Johnston}, S.,
  {Lee}, H., {Weltevrede}, P. and {Kramer}, M., 2011, \mnras, 415, 251
\bibitem[{{Lin}}(2016)]{Lin16} {Lin}, L.~C.~C., 2016, JASS, 33, 147
\bibitem[{{Lin} {et~al.}(2013){Lin}, {Hui}, {Hu}, {Wu}, {Huang}, {Trepl},
    {Takata}, {Seo}, {Wang}, {Chou}, \& {Cheng}}]{Lin&Hui}
  {Lin}, L.~C.~C., {et~al.} 2013, \apj Letter, 770, L9
\bibitem[{Kramer} {et~al.}(2006)]{kra06}
  {Kramer}, M., {Lyne}, A.~G., {O'Brien}, J.~T., {Jordan}, C.~A. and
  {Lorimer}, D.~R., 2006, Science, 312, 549
\bibitem[{Lyne} {et~al.}(2010)]{lyn10}
  {Lyne}, A., {Hobbs}, G., {Kramer}, M., {Stairs}, I. and
  {Stappers}, B., 2010, Science, 329, 408
\bibitem[{Lyne} {et~al.}(2013)]{lyn13}
  {Lyne}, A., {Graham-Smith}, F., {Weltevrede}, P., {Jordan}, C.,
  {Stappers}, B., {Bassa}, C. and {Kramer}, M., 2013, Sci, 342, 598
\bibitem[{Lyne} {et~al.}(2015)]{lyn15}
  {Lyne}, A.~G., {Jordan}, C.~A., {Graham-Smith}, F.,
  {Espinoza}, C.~M., {Stappers}, B.~W. and {Weltevrede}, P., 2015, \mnras, 446, 857
\bibitem[{Majid} {et~al.}(2017)]{maj17}
  {Majid}, W.~A., {Pearlman}, A.~B., {Dobreva}, T., {Horiuchi}, S.,  {Kocz}, J.,
  et al., 2017, \apj Letter, 834, 2
\bibitem[{Mereghetti} {et~al.}(2016)]{mer16}
  {Mereghetti}, S., {Kuiper}, L., {Tiengo}, A., {Hessels}, J.,
  {Hermsen}, W., et al., 2016, \apj, 831, 21
\bibitem[{Ng} {et~al.}(2016)]{ng16}
  {Ng}, C.~W., {Takata}, J. and {Cheng}, K.~S., \apj, 2016, 825, 18
\bibitem[{{Ray} {et~al.}(2011){Ray}, {Kerr}, {Parent}, {Abdo}, {Guillemot},
    {Ransom}, {Rea}, {Wolff}, {Makeev}, {Roberts}, {Camilo}, {Dormody}, {Freire},
    {Grove}, {Gwon}, {Harding}, {Johnston}, {Keith}, {Kramer}, {Michelson},
    {Romani}, {Saz Parkinson}, {Thompson}, {Weltevrede}, {Wood}, \&
    {Ziegler}}]{Ray2011}
  {Ray}, P.~S., {et~al.} 2011, \apjs, 194, 17
\bibitem[{Romani and Yadigaroglu,} (1995)]{ro95}
  {Romani}, R.~W. and {Yadigaroglu}, I.-A., 1995, ApJ, 438, 314

\bibitem[{Ruderman} (1969)]{rud69}
  {Ruderman}, M., 1969, \nat, 223, 597
\bibitem[{Ruderman and Sutherland} (1975)]{rud75}
  {Ruderman}, M.~A. and {Sutherland}, P.~G., 1975, \apj, 196, 51
\bibitem[{Ruderman} {et~al.} (1998)]{rud98}
  {Ruderman}, M., {Zhu}, T. and {Chen}, K., 1998, \apj, 502, 1027
\bibitem [{Spitkovsky} (2006)]{spi06}
  {Spitkovsky}, A., 2006, \apj Letter, 648, 51
\bibitem [{Takata} {et~al.}(2016)]{tak16}
  {Takata}, J., {Ng}, C.~W. and {Cheng}, K.~S., 2016, \mnras, 455, 4249
\bibitem [{Tam} (2016)]{tam16}
  {Tam}, P.~H.~T., 2016, The Astronomer's Telegram, 9365
\bibitem [{Tang} {et~al.}(2008)]{tang08}
  {Tang}, A.~P.~S. and {Takata}, J. and {Jia}, J.~J. and {Cheng}, K.~S., 2008, ApJ, 676, 652
\bibitem [{Timokhin and Harding} (2015)]{tim15}
  {Timokhin}, A.~N. and {Harding}, A.~K., 2015, \apj, 810, 144
\bibitem [{Trepl} {et~al.}(2010)]{trepl10}
  {Trepl}, L, {Hui}, C.~Y., {Cheng}, K.~S., {Takata}, J.,
  {Wang}, Y., {Liu}, Z.~Y. and {Wang}, N., 2010, \mnras, 405, 1339
\bibitem[{Wada and Shibata} (2011)]{wad11}
  {Wada}, T. and {Shibata}, S., 2011, \mnras, 418, 612
\bibitem [{Wang} {et~al.}(2010)]{wang10}
  {Wang}, N., {Manchester}, R.~N., {Pace}, R.~T., {Bailes}, M.,
  {Kaspi}, V.~M., {Stappers}, B.~W. and {Lyne}, A.~G., 2000, \mnras, 317, 843
\bibitem[{Weltevrede} {et~al.}(2011)]{wel11}
  {Weltevrede}, P., {Johnston}, S. and {Espinoza}, C.~M., 2011, \mnras, 411, 1917
\bibitem[{{Weisskopf} {et~al.}(2011){Weisskopf}, {Romani}, {Razzano},
    {Belfiore}, {Saz Parkinson}, {Ray}, {Kerr}, {Harding}, {Swartz},
    {Carrami{\~n}ana}, {Ziegler}, {Becker}, {De Luca}, {Dormody}, {Thompson},
    {Kanbach}, {Elsner}, {O'Dell}, \& {Tennant}}]{Wei2011}
  {Weisskopf}, M.~C., {et~al.}, 2011, \apj, 743, 74
\bibitem [{Wang} {et~al.}(2007)]{wan07}
  {Wang}, N.,  {Manchester}, R.~N. and {Johnston}, S., 2007, \mnras, 377, 1383
\bibitem[{Yu} {et~al.}(2013)]{yu13}
  {Yu}, M., {Manchester}, R.~N.,  {Hobbs}, G., {Johnston}, S.,  {Kaspi}, V.~M. {et~al.},
  2013, \mnras, 429, 688
\bibitem[{Yuan} {et~al.}(2010)]{yua10}
  {Yuan}, J.~P., {Wang}, N., {Manchester}, R.~N., {Liu}, Z.~Y., 2010, \mnras, 404, 280
\bibitem[{Yuki and Shibata} (2012)]{yuk12}
  {Yuki}, S. and {Shibata}, S., 2012, \pasj, 64, 43
\end{thebibliography}
\end{document}